\documentclass[12pt,preprint]{aastex}

\usepackage{epsf}
\usepackage{graphicx}
\usepackage{lscape}
\usepackage{ulem}
\usepackage{color}
\usepackage{emulateapj5}

\begin{document}
\title{A New Diagnostic of the Radial Density Structure of Be Disks}
\author{Zachary H. Draper\altaffilmark{1}, John P. Wisniewski\altaffilmark{1,2}, 
Karen S. Bjorkman\altaffilmark{3}, Xavier Haubois\altaffilmark{4}, Alex C. Carciofi\altaffilmark{4}, Jon E. Bjorkman\altaffilmark{3}, Marilyn R. Meade\altaffilmark{5}, Atsuo Okazaki\altaffilmark{6}}

\altaffiltext{1}{Department of Astronomy, University of Washington, Box 351580 Seattle, WA 98195, USA, jwisnie@u.washington.edu, zhd@uw.edu}
\altaffiltext{2}{NSF Astronomy \& Astrophysics Postdoctoral Fellow}  
\altaffiltext{3}{Ritter Observatory, Department of Physics \& Astronomy, Mail Stop 113, University of Toledo, Toledo, OH 43606, karen.bjorkman@utoledo.edu, jon@physics.utoledo.edu}
\altaffiltext{4}{Instituto de Astronomia, Geof\'isica e Ci\^encias Atmosf\'ericas, Universidade de S\~ao Paulo, Rua do Mat\~ao 1226, Cidade Universit\'aria, 05508-900, S\~ao Paulo, SP, BRAZIL, carciofi@usp.br, xhaubois@astro.iag.usp.br}
\altaffiltext{5}{Space Astronomy Lab, University of Wisconsin-Madison, 1150 University Avenue, Madison, WI 53706, meade@astro.wisc.edu}
\altaffiltext{6}{Faculty of Engineering, Hokkai-Gakuen University, Toyohira-ku, Sapporo 062-8605, Japan, okazaki@elsa.hokkai-s-u.ac.jp}

\begin{abstract} We analyze the intrinsic polarization of two classical Be stars in the process of losing their circumstellar disks via a Be to normal B star transition originally reported by Wisniewski et al.  During each of five polarimetric outbursts which interupt these disk-loss events, we find that the ratio of the polarization across the Balmer jump (BJ+/BJ-) versus the V-band polarization traces a distinct loop structure as a function of time.  Since the polarization change across the Balmer jump is a tracer of the innermost disk density whereas the V-band polarization is a tracer of the total scattering mass of the disk, we suggest such correlated loop structures in Balmer jump-V band polarization diagrams (BJV diagragms) provide a unique diagnostic of the radial distribution of mass within Be disks.  We use the 3-D Monte Carlo radiation transfer code HDUST to reproduce the observed clockwise loops simply by turning ``on/off'' the mass decretion from the disk.  We speculate that counter-clockwise loop structures we observe in BJV diagrams might be caused by the mass decretion rate changing between subsequent ``on/off'' sequences.  Applying this new diagnostic to a larger sample of Be disk systems will provide insight into the time-dependent nature of each system's stellar decretion rate.

\end{abstract}

\keywords{circumstellar matter --- stars: individual (pi Aquarii) --- stars: individual (60 Cygni)}

\section{Introduction} \label{intro}

Classical Be stars are well known to be characterized by having gaseous circumstellar decretion disks which are fed from mass-loss from their rapidly rotating central stars \citep{por03}.  As detailed in the review of \citet{car10}, the viscous decretion disk model (VDDM; \citealt{lee91}) is adept at explaining many of the observational properties of Be star disks, including the observed Keplerian rotation of the disk, and is generally considered the most promising model to explain the Be phenomenon, although alternate models also have been proposed \citep{bjo93, cas02}.  The mechanism(s) responsible for ejecting material from the stellar photosphere into a disk is also unkown, although both observational \citep{riv98,nei02} and theoretical \citep{and86,cra09} studies suggest that non-radial pulsations might act to feed at least some of these decretion disks.  

While some classical Be stars exhibit observational evidence of large-scale asymmetries \citep{oka97, vak98, ste09} interpreted as arising from one-armed density waves \citep{oka91}, the general gas disk density has typically been modeled by a very simple axisymmetric power law (see e.g. \citealt{bjo97,por99}).  Recent contemporaneous optical-infrared (IR) spectroscopic \citep{wi07a}, polarimetric \citep{car07}, and optical-IR interferometric studies \citep{tyc06,gie07, tyc08, pot10}, which are each most sensitive to different physical regions of disk, demonstrate how the radial-dependence of the gas density in these disks can be observationally constrained.  Such works offer promise for testing the appropriateness of the single power-law adopted by most models.

Some classical Be stars can exhibit stable decretion disks for decades (e.g $\zeta$ Tau; \citealt{ste09}); however, they are also known to experience aperiodic ``Be to normal B to Be'' transitions whereby they lose (and subsequently regenerate) all observational signatures of having a disk \citep{und82,cla03,vin06}.  The frequency of these transitions is not well constrained by observations, although the discovery of 12 new transient Be stars in multi-epoch study of eight open clusters suggests that these major events are not rare \citep{mcs09}.  These systems represent ideal testbeds to diagnose the fundamental mechanism which drives disk formation in Be stars precisely because  they are known to be actively losing (or gaining) a disk.  Studying such disks with techniques capable of diagnosing the radial-dependence of the gas density, especially in the inner-most regions of the disk, could enable an enhanced understanding of how these disks form.

In paper I of this series \citep{wis10}, we analyzed $\sim$15 years of spectropolarimetric observations of the classical Be stars 60 Cygni and $\pi$ Aquarii which covered one disk-loss episode in each star, and discussed the time-scale and overall evolution of these events.  In this paper, we present a first look at the behavior of the intrinsic polarization during these events and detail a unique new diagnostic which effectively traces the gas density in the inner-most region of the disk.  We investigate these ``polarization-loop'' signatures with a well vetted 3-D Monte Carlo NLTE code and present representative model runs which reproduce the observations and support our interpretation of this phenomenon.  Finally, we outline our plans to perform more detailed modeling of these signatures.

\section{Observations} \label{data}

$\pi$ Aquarii was observed 127 times between 1989 August 8 and 2004 October 10 and 60 Cygni was observed 35 times between 1992 August 3 and 
2004 September 26 with the University of Wisconsin's (UW) HPOL spectropolarimeter, mounted on UW's 0.9m 
Pine Bluff Observatory (PBO) telescope.  The full details regarding the observation, reduction, and calibration of these data, along with basic properties of the total (interstellar + intrinsic) polarization, are presented in \citet{wis10}, hereafter referred to as paper I or WIS2010.  In this paper we analyze the intrinsic polarization of these data, which were obtained by removing the interstellar polarization component described in Table 5 of WIS2010 using a modified Serkowski law \citep{ser75,wil82}.  We 
note that the residual instrumental systematic errors depends mildy on the date of the observations and ranges from 0.027-0.095\% in the U-band, 0.005-0.020\% in the V-band, and 0.007-0.022\% in the I-band.

\section{Results: A New Disk Density Diagnostic} \label{loop}

Spectropolarimetry can provide insight into the density structure of circumstellar envelopes (see e.g. \citealt{bjo00}).  Pre- or post-scattering absorption of these photons by hydrogen in the disk can imprint the wavelength-dependent opacity signature of hydrogen in the polarization if the density of absorbers is significant \citep{wo96b}, producing a ``saw-tooth'' like wavelength dependent polarization signature \citep{woo97,wi07b}.  Thus while the overall polarization is a tracer of the electron number density, or effective mass of a Be disk, the polarization change across the Balmer jump is a tracer of the largest densities of the disk, which are typically found in their innermost regions.  With these basic principles in mind, we begin our analysis of the intrinsic spectropolarimetric dataset of WIS2010.

\begin{figure}
\begin{center}
\includegraphics[width=18cm]{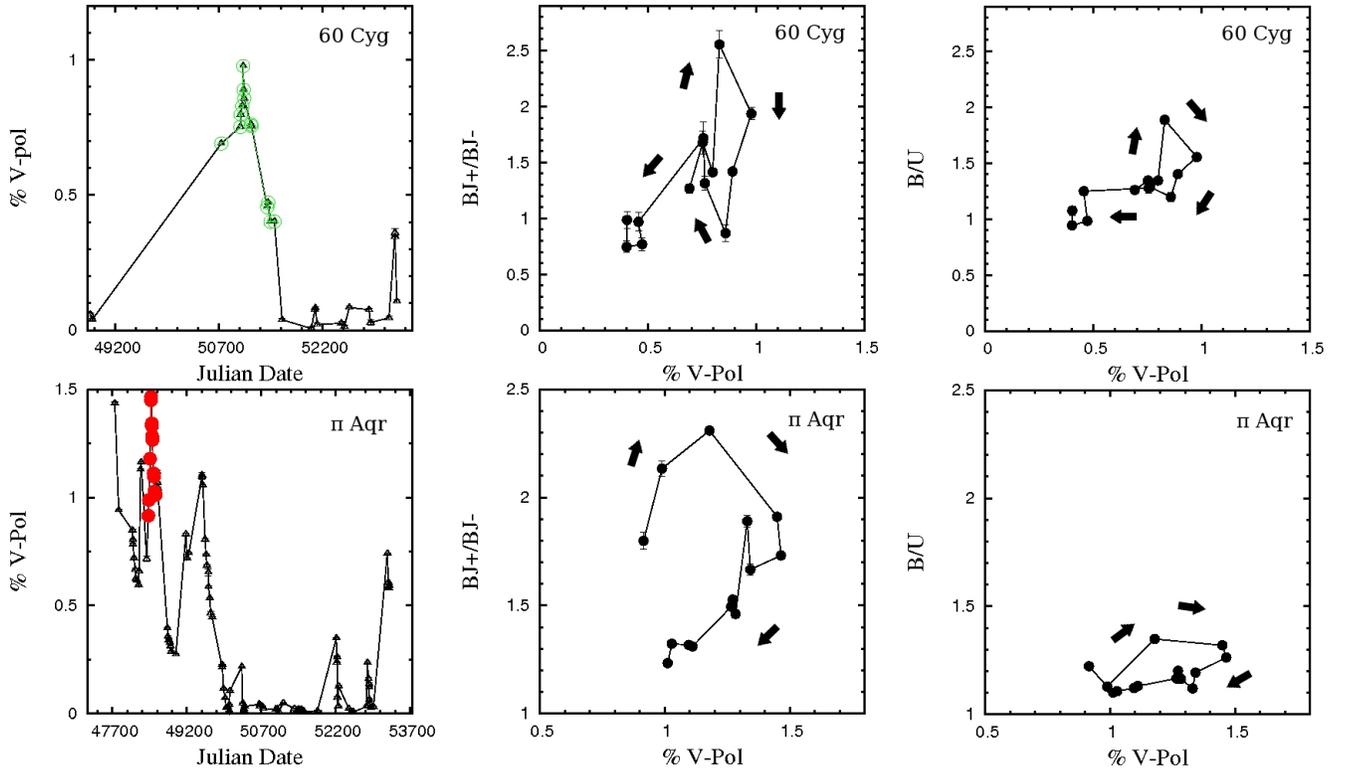}
\caption{The time evolution of the intrinsic polarization of 60 Cyg (top panels) and $\pi$ Aqr (bottom panels) is shown.  60 Cyg is characterized by a $\sim$770 day polarimetric outburst denoted by large, open green circles (top left panel).   A clock-wise loop pattern in 60 Cyg's Balmer jump-V band polarization diagram (BJV diagram; top middle panel) is observed during the early stages of the star's outbust, followed by a counter-clockwise loop during later stages of the outburst.  This loop structure is discernible at lower amplitude when the polarization ratio of the Johnson B/U filters is considered as a function of the V-band polarization (top right panel).   The time evolution of the intrinsic polarization of $\pi$ Aqr is characterized by several polarimetric outbursts, including a $\sim$180 day event denoted by large, filled red circles (bottom left panel).  A broad clock-wise loop pattern is seen in $\pi$ Aqr's BJV diagram (bottom middle panel); this loop structure is discernible at lower amplitude when the polarization ratio of the Johnson B/U filters is considered as a function of the V-band polarization (bottom right panel).  \label{loopfig} }
\end{center}
\end{figure}

WIS2010 detailed how the gradual disk-loss episodes of 60 Cyg and $\pi$ Aqr proceeded over a timescale of $\sim$1000 and $\sim$2400 days respectively, and how these events were temporarily stalled by several polarimetric outbursts (left panels Figure \ref{loopfig}; see also WIS2010).  We examined the behavior of each of these events in standard Johnson broad-band filters, created by binning our spectropolarimetric data to reproduce the coverage of each filter, and in custom filters such as one which probed the polarization across the Balmer jump.  This Balmer jump (BJ+/BJ-) filter was created simply by ratioing data binned between the wavelength range 3650 - 4100 \AA\ (BJ+) and  3200 - 3650 \AA\ (BJ-).  

During polarimetric outbursts, we detect clear evidence that the polarization across the Balmer jump (BJ+/BJ-) exhibits a distinctive loop-like evolution when compared to the V-band polarization.  We show examples of such loop-like structures in Balmer jump-V band polarization diagrams (hereafter BJV diagram) for both 60 Cyg (Figure \ref{loopfig}, top middle panel) and $\pi$ Aqr (Figure \ref{loopfig}, bottom middle panel).  The outburst responsible for each of these loops are depicted in the top and bottom left panels of Figure \ref{loopfig}.  The duration of the depicted outburst in $\pi$ Aqr is $\sim$180 days and the evolution of the polarization across the Balmer jump traces a clockwise loop pattern as a function of V-band polarization.  The duration of the depicted outburst in 60 Cyg is $\sim$770 days and while the polarization traces a clockwise loop pattern during the initial stages of the outburst, it traces a partial counter-clockwise loop pattern during the latter stages of the outburst.  We find similar clockwise loop morphologies for one additional polarimetric outburst in $\pi$ Aqr which lasted for $\sim$160 days, and a counter-clockwise loop morphology for another $\pi$ Aqr outburst which lasted for $\sim$730 days.

While these loops are very prominent in the BJV diagram (middle panels; Figure \ref{loopfig}), we also explored whether these morphological features would be detectable if one simply had broad-band filters at their disposal.  We therefore binned our spectropolarimetric data to reproduce the coverage of the Johnson U- and B-filters, and plotted our data on a (B-filter/U-filter) versus V-filter (hereafter BUV) diagram.  As seen in the right panels of Figure \ref{loopfig}, the same looplike structures are visible in BUV diagrams (as BJV diagrams), albeit at lower amplitude.  We therefore suggest that moderately high precision filter polarimetry could also provide the same effective diagnostic as afforded by our spectropolarimetric data.   

\section{Discussion} 

\subsection{Modeling the Observed Phenomenon} \label{model}

As mentioned in the introduction, the viscous decretion disk model (VDDM) is currently the most promising candidate to explain the structure, formation and evolution of Be disks. While most theoretical approaches so far considered the case of a constant mass decretion rate in the quasi-steady state limit \citep[e.g.,][]{bjo97,bjo05}, more recent studies investigated the temporal evolution of the disk surface density fed by an arbitrary mass decretion history \citep{oka07,jon08}.  Here we qualitatively investigate whether the VDDM can account for the trends we see in BJV diagrams.

In order to theoretically reproduce the observed trends shown in Figure 1 with the VDDM model we used a 1-D hydrodynamical code \citep{oka07} to compute the time-dependent surface density of the disk.
This code solves the viscous diffusion problem given a prescription for the stellar mass decretion rate, and a value for the disk kinematic viscosity (the $\alpha$ parameter of \citealt{sha73}).
The surface density for chosen epochs of the disk evolution is then fed to our radiative transfer code {\sc hdust} \citep{car06} that is capable of turning the structural information thus provided into astrophysical observables, such as emergent spectrum or intensity maps on the sky. 

\begin{figure}
\begin{center}
\includegraphics[width=15cm]{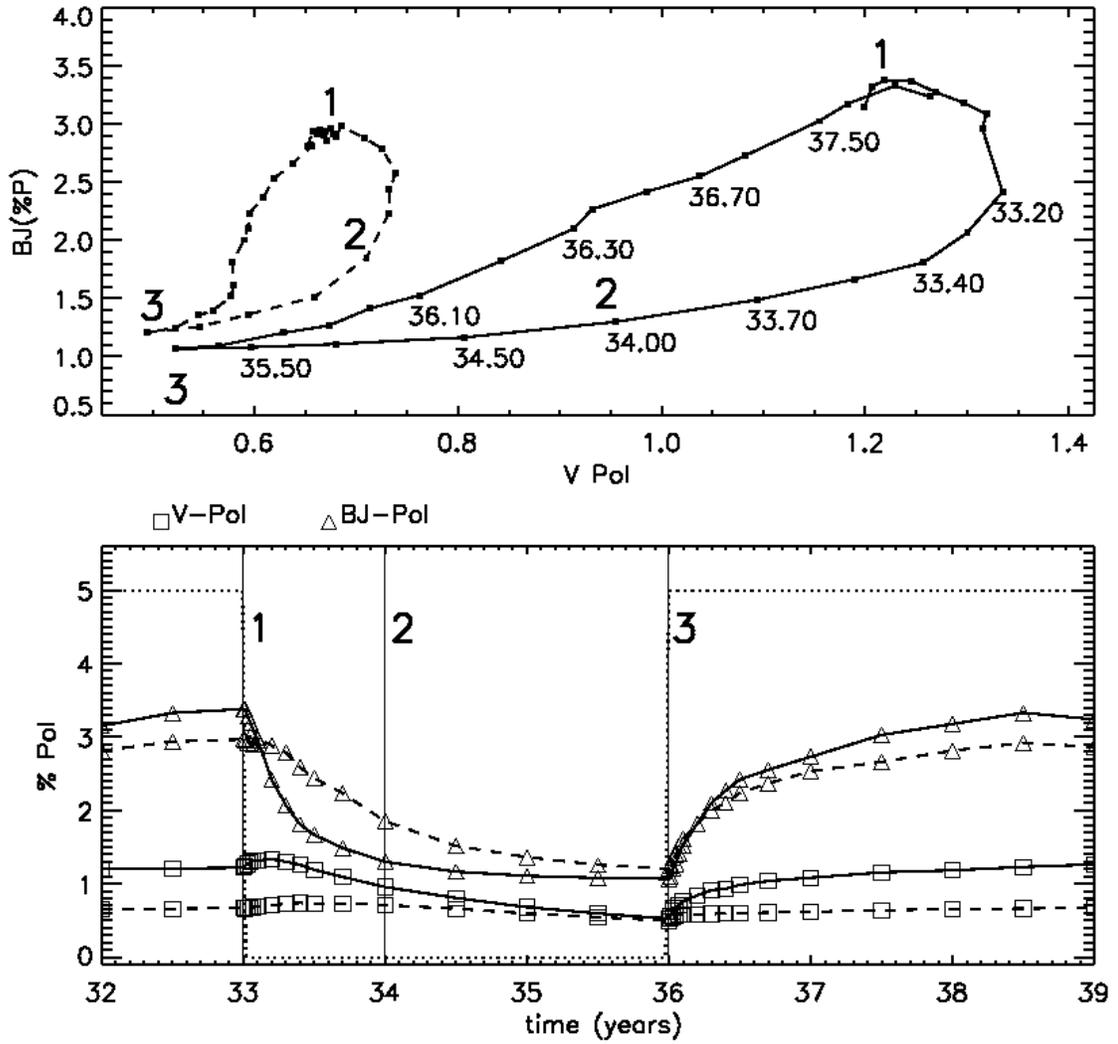}
\caption{Upper panel: BJV diagram at two inclination angles (solid line, 70 degrees and dashed line, 90 degrees) for the mass decretion rate described in Sect. 4.1.  The corresponding phase number is reported in both panels and observing epochs are indicated and counted in years.  Lower panel: Temporal evolution of the V-band polarization and polarization across the Balmer jump for both inclination angles.   The dotted line shows the mass decretion history arbitrarily scaled to the range of the graph.  \label{modelfig} }
\end{center}
\end{figure}

The correlation loops observed in the BJV diagram can be qualitatively reproduced assuming a prescription for the mass decretion rate that involves alternating cycles of activity (mass decretion on) and quiescence (mass decretion off).
One such an example is shown Fig.~\ref{modelfig}. Starting from no preexisting disk, this sample model  assumes a 3-year long period of activity followed by a 3-year long quiescence. This 6-year cycle were repeated many times. In Fig.~\ref{modelfig} we show results covering the period between 32 and 39 years after the beginning of disk formation.
The polarization forms a clockwise loop in the BJV diagram that can be described as follows. At the end of the active phase, the star had built a large and dense circumstellar disk (phase 1).
When the mass decretion stops, the inner disk quickly reaccretes back onto the star; this causes a fast drop of BJ size and the curve follows a track towards the bottom-left of the BJV diagram (phase 2). 
What follows is a slow secular dissipation of the entire disk along which the BJ size changes little (the inner disk having already reached very low densities) but the V-band polarization diminishes as the disk mass decreases (phase 2 to 3).
When the next cycle of activity begins (phase 3) the inner disk quickly fills up again and the curve  eventually reaches back the top of the BJV diagram.

The detailed shape of the loop depends on several factors: the viewing angle (Fig.~\ref{modelfig}), the value of $\alpha$, and the mass decretion history assumed. 
In the simple model shown here the loops nearly close, because the mass decretion rate assumed for each cycle is the same. The loop would not close if the mass decretion rate of subsequent cycles were to be different and/or if the length of the active/quiescent phases were irregular.  We plan to systematically explore a large range of mass decretion rate scenarios in a future publication (Haubois et al 2011, in prep).

\subsection{Comparison with CMD loops}

\citet{dew06} studied the photometric variability of several hundred Be stars in the Magellanic Clouds and found $\sim$100 stars whose photometric variations traced loop-like patterns in optical color magnitude diagrams (CMDs).  Most ($\sim$90\%) objects traced clockwise loops in these CMDs while $\sim$10\% traced counter-clockwise loops.  \citet{dew06} suggested that clockwise loops were indicative of systems actively decreting material from their stellar surfaces whereas counter-clockwise loops were indicative of systems in which material was being re-accreted onto the central star.

We note that our models of the loop-like structures we observe in BJV diagrams include the effects of both decretion of material from the central star and the re-accretion of material onto the central star.  Hence the large-scale morphological differences we observe, i.e. clockwise versus counter-clockwise loops in BJV diagrams, can not be simply attributed to differences in the radial direction of motion of material in the disk as invoked by \citet{dew06}.  Rather, we speculate that the counter-clockwise loops we sometimes observe in BJV diagrams might be caused by a non-constant mass decretion rate or two nearby cycles which have significantly different mass decretion rates.  Our future systematic exploration of the parameter space of our models (Haubois et al 2011, in prep) will enable us to test this speculative hypothesis, as well as other mass ejection scenarios, to explain these counter-clockwise loops.

\subsection{Implications and Future Applications of the Technique}

The steady-state \citep{bjo05} and time dependent \citep{oka07} surface density of Be disks is driven in large part by the ratio of the stellar
mass-loss rate and the $\alpha$ parameter (both quantities setting the disk decretion rate), although recent modeling work has demonstrated
that the disk temperature can also influence the surface density, especially in the inner disk regions, when the stellar mass-loss rate is large \citep{car08}.  As
\citet{car09} noted, observationally constraining the stellar mass-loss rate is very challenging for systems whose disks are truncated by binary
companions; moreover, constraining the detailed time dependent decretion rate of non-steady state disk
systems (e.g. \citealt{riv98}) is also challenging.  The BJV (and broad-band BUV) diagnostic we have introduced in this paper offers one
clear way to better diagnose the detailed time-dependent decretion rate of Be disk systems.
Application of this technique, when the requisite low-resolution blue optical spectropolarimetry (BJV diagram) or U- and B-band filter
polarimetry is available, to a larger sample of Be systems actively gaining/losing their disks would help elucidate the mechanism(s)
responsible for triggering disk formation in Be stars.  We encourage the community to obtain this type of well time sampled polarimetry for the 
mid-2011 periastron passage of $\delta$ Sco, as it would be provide a powerful diagnostic of the type of decretion outbursts which have been reported 
in previous periastron passages \citep{mir01,mir03}.  Moreover, inclusion of simple V-band photometry in such analysis could help
observationally determine the $\alpha$ parameter \citep{car10}.

\section{Summary}

We have analyzed the intrinsic polarization of 60 Cyg and $\pi$ Aqr as they were in the process of losing their circumstellar disks via a Be to normal B star transition.  During each of five polarimetric outbursts which interupt these disk-loss events, we find that the ratio of the polarization across the Balmer jump (BJ+/BJ-) versus the V-band polarization traced distinct loop structures as a function of time, and suggest that this diagnostic provides unique insight into the radial distribution of mass within Be disks.  We observed two clockwise and one counter-clockwise loop structures in Balmer jump-V band polarization diagrams (BJV diagragms) of $\pi$ Aqr, while 60 Cyg exhibited a combined clockwise and counterclockwise loop (i.e. a ``figure eight'').  We use the 3-D Monte Carlo radiation transfer code HDUST to reproduce the observed clockwise loops simply by turning ``on/off'' the mass decretion from the disk, and speculate that counter-clockwise loop structures might be caused by the mass decretion rate changing between subsequent ``on/off'' sequences.  Current and future exploration of the parameter space of our models (\citealt{hau10}; Haubois et al 2011 in prep) will help efforts to identify the definitive origin of these counter-clowckwise loop structures.

\acknowledgments

We thank Brian Babler for his assistance with HPOL data, and Kenneth H. Nordsieck for providing access to the HPOL spectropolarimeter.  We also thank our referee, David Harrington, for his suggestions which improved the clarity and content of this paper.  JPW acknowledges support from NSF Astronomy \& Astrophysics Postdoctoral Fellowship AST 08-02230 and a Chretien International Research Grant.  
ZHD was supported by the University of Washington Pre-MAP program, a UW College of Arts and Sciences Research Scholarship, and a NSF REU at the University of Toledo. 
ACC acknowledges support from CNPq (grant 308985/2009-5). XH acknowledges support from Fapesp (grant 2009/07477-1).
HPOL observations were supported under NASA contract NAS5-26777 with the University of Wisconsin-Madison.

\newpage
\clearpage
 
\newpage
\clearpage
 
\end{document}